\documentclass{ws-procs9x6}
\usepackage{amssymb}

\begin{document}

%%%%%%%%%%%%%%%%%%%%%%%%%%%%%%%%%%%%%%%%%%%%

%%%%%%%%%%%%%%%%%%%%%%%%%%%%%%%%%%%%%%%%%%%%%

\title{QUARK COUNTING RULES: \\  OLD AND NEW APPROACHES\footnote{Talk  given at ``Shifmania'',
Minneapolis, MN, May 14-17, 2009}}

\author{A.  RADYUSHKIN\footnote{Also  at Bogoliubov Laboratory of Theoretical Physics, JINR, Dubna, Russian Federation}}

\address{  Physics Department, Old Dominion University, 
 Norfolk, VA 23529, USA \\ 
 and  \\
 Theory Center, Jefferson Lab,  Newport News, VA 23606, USA
}

\begin{abstract}
I  discuss the subject  of powerlike  asymptotic behavior of hadronic form factors 
in pre-QCD analyses of  soft (Feynman/Drell-Yan)  and hard
(West)  mechanisms, and also recent  
derivation of $1/Q^2$ asymptotics of meson form factors in AdS/QCD.  
At the end, I briefly comment on ``light-front  holography'' ansatz.
\end{abstract}

\bodymatter

 \section{Hadronic form factors} 
 
{\it Introduction.}  Experimental evidence  that (exclusive) form factors of hadrons
consisting of $n_q$ quarks  behave  like  $(1/Q^2)^{n_q-1}$ for large $Q^2$,
provokes  expectations  that there is a fundamental and/or easily visible 
reason for such a phenomenon, scale invariance being   the 
 most natural suspect\cite{Matveev:1977ra}. 
 Indeed,  hard rescattering  in a theory with spinor constituents and 
dimensionless coupling constant for   their interaction with an intermediary  boson field 
provides a 
specific dynamical mechanism\cite{Brodsky:1973kr} that produces the $(1/Q^2)^{n_q-1}$
  behavior.  In this approach, $n_q-1$ is  just the number of hard exchanges. 
Another property  apparently correlated with the number of quarks in the hadron
is the $\sim (1-x)^{2n_q-3}$ behavior of the (inclusive) 
 quark  distributions functions  in the $x \to 1$ region.  
This observation suggests to look for connection between these  exclusive 
and inclusive observables.
Below  in this section  we discuss   scenarios  which display   two  versions
of exclusive-inclusive correlation.
In subsequent sections, we  discuss derivation of the $1/Q^2$ behavior 
for meson form  factors in AdS/QCD.

{\it Soft mechanism.} 
Powerlike  behavior of hadronic form factors due to   
Feynman mechanism  \cite{Feynman:1973xc}  can  be derived from the 
 Drell-Yan formula  \cite{Drell:1969km}
\begin{align}
F(Q^2) =  \int_0^1 dx \int d^2 {\bf k}_\perp 
\,  \Psi^* (x,{\bf k}_\perp+(1-x)  {\bf q}_\perp )
 \Psi  (x,{\bf k}_\perp)  \  , 
\label{DY} 
\end{align}
which represents form factor in terms of the light-front 
wave function  $\Psi  (x,{\bf k}_\perp)$  and light-front
variables $x$ and ${\bf k}_\perp$.
When  the wave   function 
$\Psi  (x,{\bf k}_\perp)$ rapidly  (say, exponentially) 
decreases  for $k_\perp\gtrsim \Lambda$,  it is natural  to
consider    the region where  both   $ \Psi  (x,{\bf k}_\perp)$
and $\Psi_M^* (x,{\bf k}_\perp+\bar x {\bf q}_\perp )$ are maximal: 
{$i)$} $ |{\bf k}_\perp \lesssim \Lambda $  is small      and  
{$ii)$}   $\bar x \equiv 1-x$ is close to 0, 
so that  $|\bar x {\bf q}_\perp|
\lesssim \Lambda $.  
If $| \Psi  (x, {\bf k}_\perp \lesssim  \Lambda)|^2 \sim (1-x)^{2n_q-3}$  then
\begin{align}
 F(Q^2) \sim \int_0^{\Lambda/Q}\bar x^{2n-3} \,  
d \bar x \sim  (1/Q^2)^{n_q-1} \  .  
\end{align}
The parton  distribution  functions  in this formalism are given by the 
integral of $ |\Psi  (x,{\bf k}_\perp)|^2$ over ${\bf k}_\perp$. 
The latter is  dominated by $k_\perp \lesssim \Lambda$, 
hence  \mbox{$f(x)  \sim (1-x)^{2n_q-3}$.}  Thus, changing the shape of $f(x)$,
one would  change the result for  form factor. 
In  other words, there is a  {\it causal relation} between the $x\to 1$ shape of the
distribution function $f(x)$ and the $Q^2 \to 1$ behavior of the form factor $F(Q^2)$: 
form of $f(x)$ determines $F(Q^2)$.

 {\it Hard  mechanism.}
For the Feynman/DY mechanism it was important that the fraction $\bar x  \equiv 1-x$ 
vanishes in the $Q^2 \to 0$ limit. 
Consider now  the regions in DY formula (\ref{DY}), in which   
the  fraction  $\bar x$ is  finite,  while  the transverse momentum argument 
of one  of the wave functions is  small,   e.g.,  the region 
$|{\bf k}_\perp | \ll \bar x  |{\bf q}_\perp|$,  
where  $ \Psi (x,{\bf k}_\perp)$ 
 is maximal. Then 
\begin{align}
 F (Q^2) 
\sim    
 \int_0^1 |  \Psi^* (x, \bar x {\bf q}_\perp ) \, 
\varphi(x)  |  \,  {dx} \  , 
\label{DYhard} 
\end{align}
where 
\begin{align}
 \varphi(x)  =  \int 
\,  
 \Psi  (x,{\bf k}_\perp) \, d^2 {\bf k}_\perp 
\label{DA} 
\end{align}
is the relevant distribution amplitude.  
In this   scenario, the  form factor repeats large-${\bf k}_\perp$
behavior of the hadron wave function, e.g., if $ \Psi  (x,{\bf k}_\perp) \sim (1/{\bf k}_\perp^2)^{n}$, 
then $ F (Q^2)  \sim (1/Q^2)^{n}$. 
This mechanism  was proposed by G.B. West \cite{West:1970av}, who  used, in fact,  
a  covariant Bethe-Salpeter (BS)   formalism rather than light-front 
variables, writing   the form factor as\cite{West:1970av} 
\begin{align}
F(Q^2) \sim  \int  f(p) \, f(p+q)\,  d^4 p  \  , 
\end{align}
\begin{figure}[htb]
  \centerline{ \includegraphics[height=3cm]{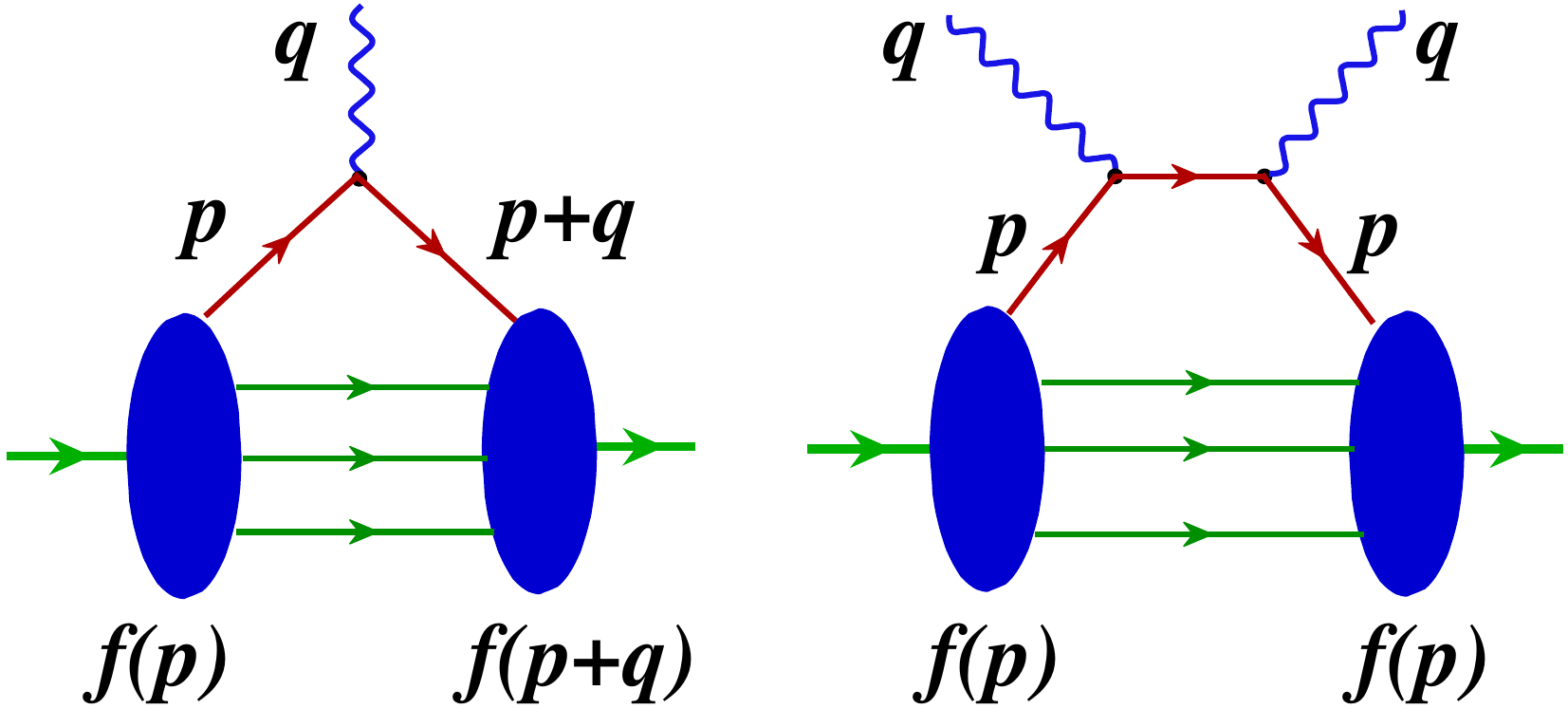}}
\caption{Form  factors and structure functions  in West's model.} 
\end{figure}
where $f(p)$  is  treated as   a function of the active parton virtuality 
$t\equiv p^2$   and spectator  
mass $M^2$. 
Assuming that  $f(t, M^2) \sim t^{-n} g(M^2)$ for large $t$, 
West  concludes that  $ F(Q^2) \sim (1/Q^2)^n$:
form  factor repeats the large-$Q^2$ behavior of the BS wave function $f(p+q)$. 
For the deep inelastic structure function, West obtains\cite{West:1970av}  
\begin{align}\nu W_2 (x) \sim \int_{t_{\rm min}}^{t_{\rm max}\sim -2 \nu}
f^2(t,M^2)  \, dt \sim ({t_{\rm min}})^{-2n+1}  \  , 
\end{align} 
where $t_{\rm min} = \left (\frac{-x}{1-x}\right )\left [M^2 -(1-x)M_N^2\right    ]$, $M_N$ being the nucleon mass.  
As a result, 
$
 \nu W_2 (x) \sim (1-x)^{2n-1} 
$.

{ \it DY  vs West's  model.}  If $n=n_q-1$, the power-law predictions of 
the two  models  formally  coincide.
However,  these  results were obtained from     completely  different  assumptions.
In  DY picture, the 
active parton is ``on-shell''  both before and after the collision:
both $|{\bf k}_\perp| $  and $|{\bf k}_\perp+ \bar x {\bf q}_\perp| $ are of order  $ \Lambda$,
and  form factor 
$F(Q^2)$ reflects the  {\it size of  phase space}  region     in which $1-x \sim \Lambda/Q$ . 
On the other hand, in   West's  model, the active parton is highly virtual
either in initial or final state, and 
$F(Q^2)$ reflects the $t$-dependence  of  WF for large virtualities $t=p^2$. 
Still, though the two mechanisms are completely different,  
the connection \mbox{$(1/Q^2)^{n}\Leftrightarrow (1-x)^{2n-1}$} (``Drell-Yan-West relation'') 
holds in  both  models!\footnote{This  is apparently 
 why the two models are confused  up 
to the point that Eq.~(\ref{DY}) is often  referred to as ``Drell-Yan-West  formula'',
which  is  absolutely incorrect because its  crucial feature is incorporation of 
light-front variables that  West did not use.}  
It should be also emphasized that in West's model,  
$(1/Q^2)^{n} $ and $ (1-x)^{2n-1}$
have the same cause (large-$t$ behavior of $f(p)$), but they are not ``causing'' each  other.

{\it West's hard  mechanism   \& pQCD.}
In DY model, $n$ is  not  necessarily  integer. 
Integer   values of  $n$ naturally  appear in West's 
hard scenario, where  they are  related to  the   number of hard 
propagators. 
In particular,  hard exchange  in a theory with a 
 dimensionless  coupling constant   gives $n=n_q -1$ [\refcite{Brodsky:1973kr}], which  is 
 a consequence   of scale invariance\cite{Matveev:1977ra}.  
In  quantum chromodynamics, each hard gluon exchange 
is accompanied by  effective coupling constant $\alpha_s$, i.e., 
 $F_{n_q} (Q^2) \sim (\alpha_s/Q^2)^{n_q-1}$. 
According to explicit calculation\cite{Radyushkin:1977gp,Lepage:1979zb}, 
the asymptotic prediction
 for the pion form factor in pQCD is 
 $F_\pi (Q^2) \to  (2\alpha_s/\pi)\,s_0/Q^2$, where 
 $  s_0 = 4 \pi^2 f_\pi^2 \approx 0.7{\rm \,GeV}^2  \sim  m_{\rho}^2 $.
Compared to the VMD expectation $F_\pi (Q^2) \sim m_{\rho}^2/Q^2$, pQCD prediction is suppressed 
by $2\alpha_s/\pi$ factor. 
It is  well known that 
the factor  $\alpha_s/\pi \sim 0.1$  is penalty for an extra loop, 
which suggests that the hard  one-gluon-exchange contribution 
is an  ${\cal O} (\alpha_s)$  correction to some  ${\cal O} (\alpha_s^0)$   term.
The only candidate is the Feynman/DY 
soft contribution, which should be calculated in a  nonperturbative way.
In particular,   in  holographic 
AdS/QCD models considered in  Refs.~[\refcite{Grigoryan:2007wn,Grigoryan:2008cc}]  
one has $F_\pi (Q^2) \to  s_0/Q^2$,
without a  suppression factor.

\section{Vector meson form factors in  AdS/QCD}

Models   based on AdS/CFT  correspondence are 
  often  claimed to provide    nonperturbative explanation  
of  quark counting rules for   form factors that is  
 based on conformal invariance and   short-distance 
  behavior of 
 normalizable modes $\Phi (\zeta)$ playing the role of wave functions 
of initial and final hadrons.  Namely, in the model of 
Polchinski and Strassler\cite{Polchinski:2001tt,Polchinski:2002jw} 
(that involves on the AdS side scalar fields only) one has\cite{Brodsky:2006uqa}
\begin{align}
F(Q^2)
= \int_0^{1/\Lambda} 
 \Phi_{P'}(z) J(Q,z) \Phi_P(z) \, {dz}/z^{3}   \  , 
\label{AdSFF}
\end{align} 
where $J(Q,z) = 
z Q K_1 (z Q)\equiv {\cal K}_1 (z Q)$ is  
nonnormalizable mode describing the probing EM current, and normalizable modes for mesons
are given by 
$\Phi(z) = C z^2 J_{L+1}(\beta_{L,k} z \Lambda)$,
with $K_1$ and $J_{L+1}$ being standard Bessel functions. 
For large $Q$, one  may approximate  ${\cal K}_1 (z Q) \sim e^{-z Q}$,
and it is  clear that 
only small $z \lesssim  1/Q$  contribute. As a result, 
\mbox{$F_{L=0}  (Q^2) \to 1/Q^4$}   for the ground state.  
But this is  not the $1/Q^2$ power that one is longing   to get! 
To  bring the result  of this AdS/CFT-based model in agreement 
with pQCD expectations, Brodsky and de Teramond proposed\cite{Brodsky:2006uqa}   
to modify the basic principle of AdS/CFT correspondence,
requiring that   the dimension of the operator on the AdS side  should be equal to the
twist of the corresponding current   in the 4-dimensional theory 
rather than  to  its dimension.  
In our papers with  H.R. Grigoryan  \cite{Grigoryan:2007vg,Grigoryan:2007my,Grigoryan:2007wn,Grigoryan:2008up,Grigoryan:2008cc} we  
demonstrated that 
in more realistic  AdS/QCD models of Refs.\cite{Erlich:2005qh,Karch:2006pv}
it is possible to get 
$F_{L=0}  (Q^2) \to 1/Q^2$ for  (leading) 
meson form factors without challenging  
 the  Maldacena\cite{Maldacena:1997re}  
 correspondence
principle.

{\it Hard-wall  model}   is  formulated in 
5-dimensional space $\{ x^\mu, z\}\equiv X^M$
having 
 AdS$_5$ metric 
$
ds^2 =\left(\eta_{\mu \nu}dx^{\mu}dx^{\nu} -
dz^2\right)/z^2
$
with a hard wall:   \mbox{$ 0 \leq z \leq z_0 = 1/\Lambda$.  }
The  basic object is the 
5-dimensional  (5D) vector gauge field $A_{M} (X)$ ($M={\mu, z}$) which produces  
   4D  field  
$ 
A_{\mu}(x)=  A_{\mu}(x,z=0) 
$.
at the UV boundary of AdS space. 
The 5D gauge action   for the vector   field is  given by 
\begin{align}
 S_{\rm AdS} = - \frac{1}{4g_5^2}\int d^4x~dz~\sqrt{g}~{\rm
Tr}\left(F_{MN}F^{MN}\right)  \  ,
\end{align}
 where   $F_{MN} $ is the field-strength   tensor. The 
coupling constant \mbox{$g_5^2 = 6 \pi^2 /N_c$}  is small in  large-$N_c$ limit. 
The  free field satisfies $ \Box_5 A (X) =0$ or 
\begin{align}
 \Box_4 A (x,z) +z\partial_z \left(\frac{1}{z} \,  \partial_z A (x,z) \right )=0 \ . 
\end{align}
In 4D  momentum    representation this gives 
\begin{align}
 z\partial_z \left(\frac{1}{z} \,  \partial_z
\tilde A(p,z)\right) + p^2 \tilde A (p,z) = 0  \  .
\label{EOM}
\end{align}
According to  AdS/QCD correspondence 
\begin{align}
  \tilde A_{\mu}(p,z) = \tilde{A}_{\mu}(p)\,  {V(p, z)}/{V(p,0)} \equiv  \tilde{A}_{\mu}(p)\, {\cal V}(p,z)  \  ,
\end{align}
where the 
 bulk-to-boundary  propagator $V(p, z)$  satisfies  Eq.(\ref{EOM}).
The  gauge-invariant boundary 
condition (b.c.) $ F_{\mu z}(x, z_0) = 0 $ on the infrared (IR)  wall  
results in   { Neumann} b.c.  $ \partial_z V (p, z_0) = 0 $, with solution
\begin{align}
V(p,z) = Pz\left [ Y_0(Pz_0)J_1(Pz) - J_0(Pz_0)Y_1(Pz) \right ] \  .
\end{align}
Using  Kneser-Sommerfeld  formula\cite{Kneser} 
gives   bound  state expansion
\begin{align}
& {\cal V}(p,z) = -  \sum_{n =
1}^{\infty}\frac{ g_5 f_{n} }{p^2 - M^2_{n} } \, \psi_n( z) 
\end{align}
with masses:   $ M_{n} = \gamma_{0,n}/z_0 $ determined by zeros  $J_0(\gamma_{0,n})=0  $
of Bessel functions,  while the 
 ``coupling constants''  $f_n$  are given by 
\begin{align}
  f_{n} = \frac{\sqrt{2} M_n} {g_5 z_0  J_1(\gamma_{0,n})} \  . 
\label{fn} 
  \end{align}
They are accompanied by  ``$\psi$'' wave functions
\begin{align}
 \psi_n( z)   = \frac{\sqrt{2} }{z_0 J_1(\gamma_{0,n})}\, z J_1(M_{n} 
z) 
\end{align}
coinciding with nonnormalizable modes of 
Polchinski-Strassler model \cite{Polchinski:2001tt,Polchinski:2002jw}. 
These ``$\psi$'' 
wave  functions (w.f.) 
obey  equation of
motion  (\ref{EOM})  with \mbox{$ p^2 = M^2_{n} $, } 
satisfy  $\psi_n(0) = 0$  at UV  boundary,  and  $\partial_z \psi_n(z_0) = 0$  at   IR   boundary. 
They  are normalized according to
\begin{align}
\int^{z_0}_0~  |\psi_n(z)|^2 \, \frac{dz}{z} \, = 1  \  .
\end{align}
\begin{figure}[ht]
   \centerline{ \includegraphics[height=2.3cm]{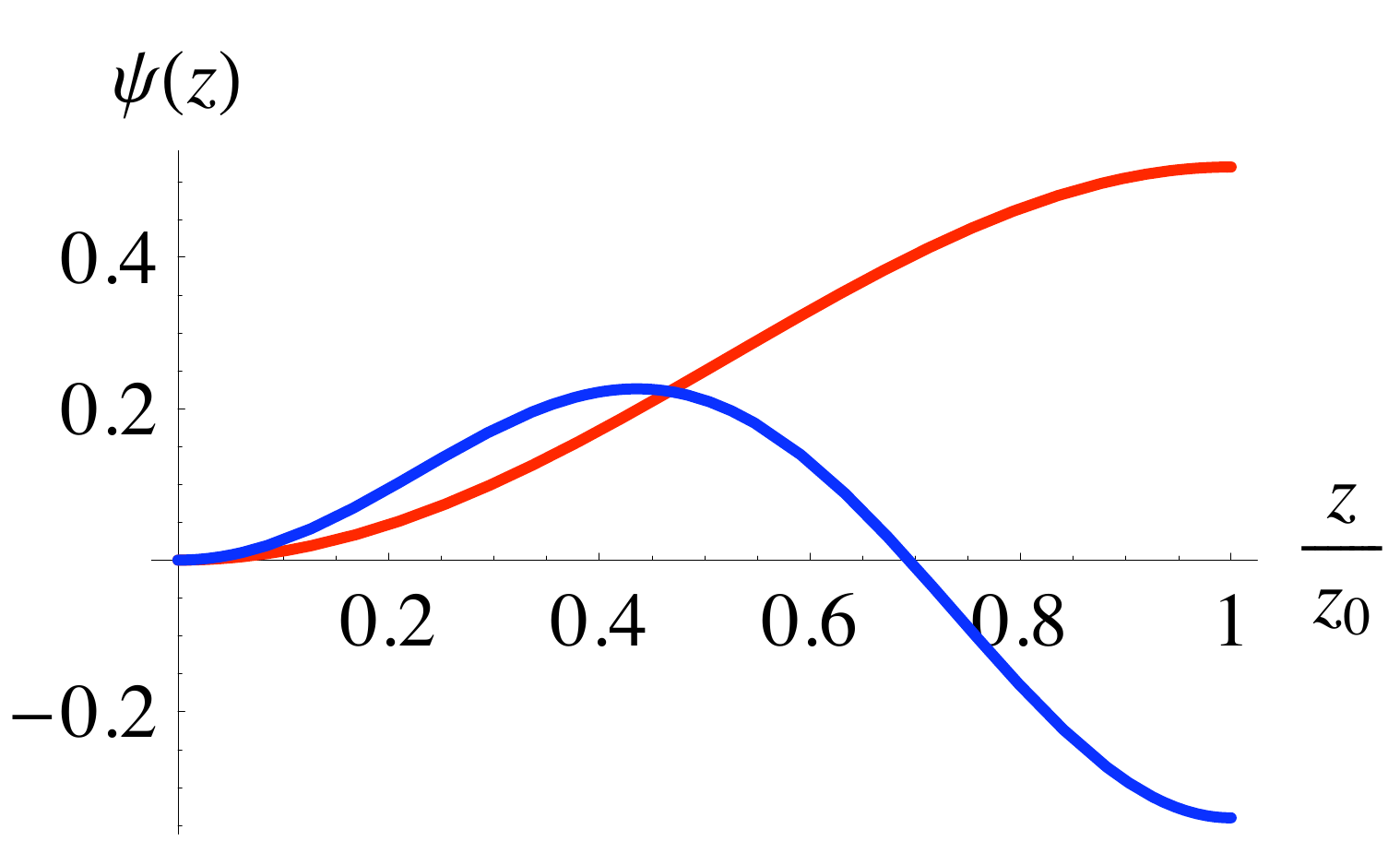}
      \  \   \includegraphics[height=2.3cm]{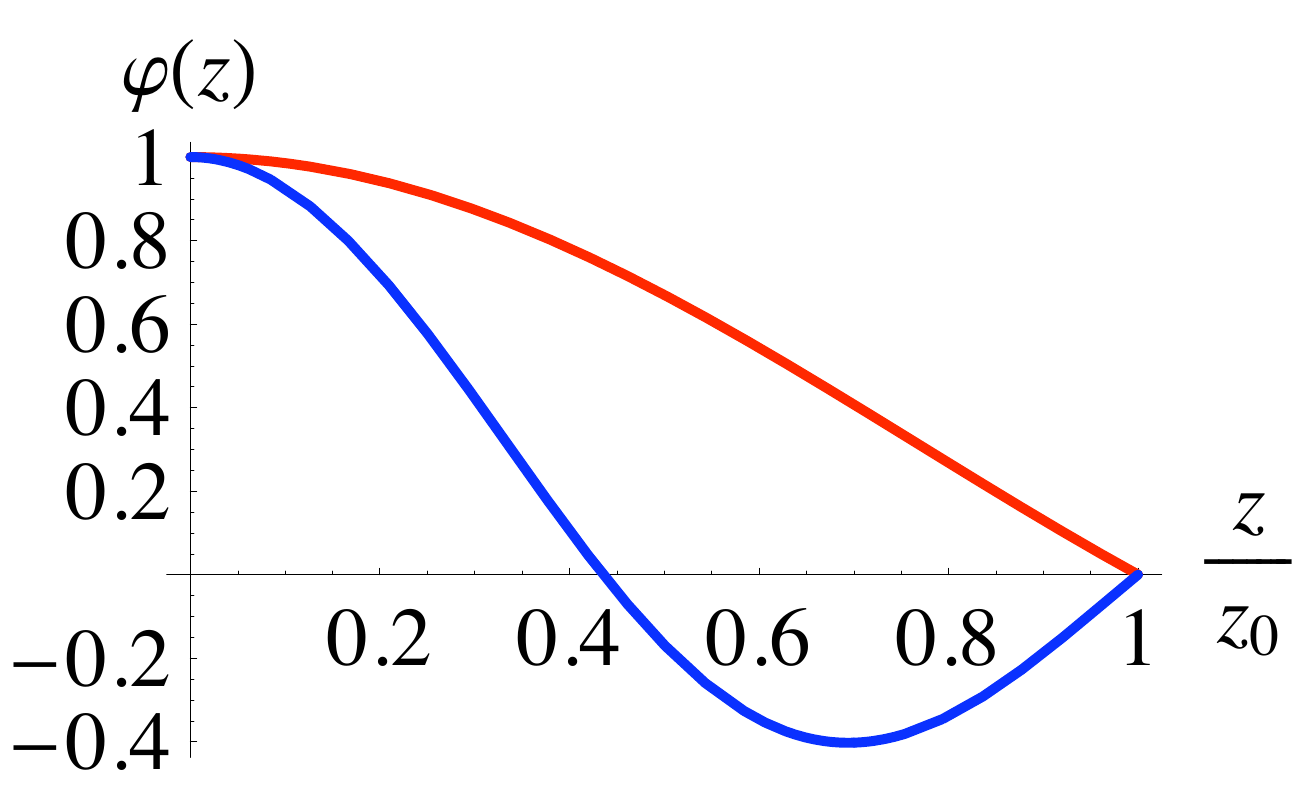}
      }
\caption{$\psi$ and $\phi$ wave functions for vector mesons.}
\label{psivec}
\end{figure}
 However, they do   not  look  like bound  state w.f. in   quantum   mechanics,
see Fig.\ref{psivec}, left. 
To this end, it  makes sense to introduce   ``$\phi$''   wave functions
\begin{align}
 \phi_n(z) \equiv  \frac1{M_n z}\, \partial_z \psi_n (z) = \frac{\sqrt{2} }{z_0  J_1(\gamma_{0,n})}\,
J_0(M_{n} z)  \  .
\end{align}
 According to  Sturm-Liouville   equation (\ref{EOM}), they are reciprocal to ``$\psi$'' w.f.:
\begin{align}
 \psi_n (z)=- {z} \  \partial_z \phi_n (z) /{M_n }  \  .
\end{align}
The  $\phi$  w.f.  give   couplings $g_5 f_n/M_n$ as   their values at   the
origin,  they 
satisfy  Dirichlet  b.  c.
$\phi_n(z_0)=0 $ at   confinement  radius, and 
 are normalized   by
\begin{align}
  \int_0^{z_0}  | \phi_n(z)|^2 z\, dz    =   1  \  .
\end{align}
The ``$\phi$'' w.f. (see Fig.\ref{psivec},right) are thus analogous   to   
bound   state wave functions  in  quantum mechanics.
The difference between the two types of AdS/QCD  wave functions  can be easily understood: 
$\psi$  w.f. correspond to vector-potential $A_M$, while 
$\phi$ w.f. correspond to field strength tensor $F_{MN}$.

  {\it Three-point function} should be introduced to study  form factors. 
It has a 
``Mercedes-Benz''  form 
\begin{align}
 W(p_1,p_2,q) = \int^{z_0}_{0}
\,{\cal V}(p_1,z) 
{\cal V} (p_2,z)  \, {\cal V} (q,z)  \, {\frac{dz}{z}} \  .
\end{align}
 For   spacelike $q$    (with $q^2=-Q^2$)  we have 
$ {\cal V} (iQ,z)  \equiv  {\cal J}(Q,z) $
The form    factors for diagonal $n \to n$  transitions may  be written 
\begin{align}
F_{nn}(Q^2) = \int^{z_0}_{0} {\cal
J} (Q,z) \, |\psi_n (z) |^2 \, \frac{dz }{z} 
\end{align}
either in  terms of $\psi$   functions or in terms of $\phi$   functions\cite{Grigoryan:2007vg}
\begin{align}
 F_{nn}(Q^2) = \frac1{1+{Q^2}/{2M_n^2}}
 \int^{z_0}_{0}   {\cal J} (Q,z) \, |\phi_n
(z) |^2 \, {z} \, {dz }  \   .
\end{align}
The overlap integral here  is a direct
analogue  of  form   factors in  quantum   mechanics, so we define 
\begin{align}
 {\cal F}_{nn}(Q^2) \equiv 
 \int^{z_0}_{0} {\cal J} (Q,z) \, |\phi_n
(z) |^2 \, z \, {dz }    \  .
\label{calF}
\end{align}
 The hard-wall model calculation gives 
\begin{align} 
& \langle \rho^{+}(p_2,\epsilon')   |J^{\mu}_{\rm EM}
(0)|\rho^{+}(p_1,\epsilon) \rangle \nonumber \\ & = 
 - \epsilon'_{\beta} \epsilon_{\alpha}
 \bigl  [\eta_{\alpha\beta}(p_1+p_2)_{\mu} + 2(\eta_{\alpha \mu}q_{\beta} -
\eta_{\beta \mu}q_{\alpha}) 
\bigr ]  F_{nn}(Q^2) \  .
\end{align}
But it is well known that  vector   mesons  have three form  factors:   
\begin{align} 
& \langle \rho^{+}(p_2,\epsilon')   |J^{\mu}_{\rm EM}
(0)|\rho^{+}(p_1,\epsilon) \rangle =  -
\epsilon'_{\beta} \epsilon_{\alpha} \bigl[ \ \eta^{ \alpha \beta}
(p_1^{\mu}  + p_2^{\mu} )\, G_1(Q^2)    \\  & +
 (\eta^{\mu \alpha}q^{\beta} - \eta^{\mu \beta}q^{\alpha} )
(G_1(Q^2) + {G}_2(Q^2) )  -  \frac{1}{M^2} q^{\alpha}
q^{\beta} (p_1^{\mu}+p_2^\mu )\,  G_3(Q^2) \ \bigr]  \  , 
 \nonumber
\end{align}
i.e.,   $G_1 (Q^2)=G_2 (Q^2) =
F_{nn}(Q^2)$ and $G_3  (Q^2) =0$ [\refcite{Son:2003et}].
The form factor (\ref{calF}) is projected by taking the ``+++''  component   of  3-point
correlator,  
\begin{align}
{\cal F} (Q^2) = G_1(Q^2) + \frac{Q^2}{2M^2}\, G_2(Q^2) - \left (
\frac{Q^2}{2M^2} \right )^2 \,G_3(Q^2) \  .
\end{align}
For $\rho$-meson, this  combination  coincides with the { IMF } 
``$LL$''  transition having $\sim \alpha_s/Q^{2}$ behavior in pQCD\cite{Brodsky:1992px}. 
Taking the hard-wall model    prediction (\ref{calF}) 
and using that  $z\sim 1/Q $  dominate in the 
large-$Q$ limit  because 
$
{\cal J} (Q,z) \to zQ K_1 (Qz)  \sim e^{-Qz} 
$,
we may   substitute $\phi (z)$ 
by $\phi (0)$. 
Thus, 
\begin{align}
{\cal F}(Q^2)  \to \frac{ |\phi (0)|^2}{Q^2} 
\int_0^\infty d\chi \, \chi^2 \,  K_1 (\chi) = 2 \,\frac{ |\phi (0)|^2}{Q^2}  \  ,
\label{largeQ}
\end{align}
and  we get the same power of $1/Q^2$  as in pQCD, but without  $\alpha_s/\pi$  factor.

{\it Soft-wall   model}\cite{Karch:2006pv} 
corresponds to  $z^2$ barrier, and 
bulk-to-boundary propagator $ {\cal V} (p,z) $     can be written
 ($a = - p^2/4\kappa^{2} $) as\cite{Grigoryan:2007my} 
\begin{align}
 {\cal V}(p,z)  =  a \int_0^1 dx  \, x^{a-1} \, \exp \left [ - \frac{x}{1-x} \, \kappa^{2}z^2  \right ] \ . 
\label{Softprop}
\end{align}
  The     propagator  poles  are   located at $p^2=4(n+1)\kappa^{2 }\equiv   M_n^2$ [\refcite{Karch:2006pv}]: 
\begin{align}
 {\cal V}(p,z)  = \kappa^{2} z^2   \sum_{n=0}^{\infty} 
\frac{L_n^1 (\kappa^{2} z^2)}{a+n+1}   =  \sum^{\infty}_{n =
0} \frac{ g_5  f_n }{ M^2_n  -  p^2}  \,  \psi_n (z)  \  . 
\end{align}
Just   like in the hard-wall case,  we  deal with $\psi$ wave   functions
and 
coupling   constants $g_5 f_n$ given by their derivatives at   the origin
\begin{align}
 g_5 f_n = \left.  
\frac{1}{z}\, e^{-\kappa^{2} z^2} \, \partial_z  \psi_n (z) 
\right|_{z = \epsilon \rightarrow 0} = \sqrt{8(n+1)}\kappa^{2} \  .
\label{fnS}
\end{align}
Again, we can introduce the (Sturm-Liouville-)  conjugate $\phi$ wave   functions: 
\begin{align}  \phi_n(z)=  \frac{1}{M_n z}e^{-\kappa^{2}z^2}\partial_z \psi_n(z) = 
\frac{2}{M_n}e^{-\kappa^{2}z^2} L^0_n(\kappa^{2}z^2)  \  . 
\end{align}
 Taking the   diagonal form   factor  for the   lowest state
\begin{align}
{\cal F}_{00} (Q^2)  = 
{2} \int^{\infty}_{0} e^{-\kappa^{2}z^2} 
{\cal J}(Q,z) \, z \, dz 
\end{align}
 and using representation (\ref{Softprop}) for ${\cal J}(Q,z) $   gives 
$
 {\cal F}_{00} (Q^2)   = 1/(1+Q^2/M_0^2) \  ,
$
 i.e.,  exact    vector meson  dominance.  
 Large-$Q^2 $   behavior of  ${\cal F}$   form   factor is  given  by the same expression
(\ref{largeQ}) 
 as in hard-wall  model, the only difference being in the value
of w.f. at the origin. As a result,  we have 
\begin{align}
  {\cal F}_{\rho}^{\rm H}
		  (Q^{2 }) \to {2.56 \, m_{\rho}^{2}}/{Q^{2}} \  \  , \  \ 
 {\cal F}_{\rho}^{\rm S}
		  (Q^{2 }) \to {m_{\rho}^{2}}/{Q^{2}} \  . 
\end{align}

\section{Pion Form Factors  in AdS/QCD}

 {\it The full action} of hard-wall model\cite{Erlich:2005qh} is  given by 
\begin{align}
 S^{B}_{\rm AdS} &= {\rm Tr}  \int d^4x \int_0^{z_0}
 dz~\biggl[\frac{1}{z^3}(D^{M}X)^{\dagger}(D_{M}X) +
\frac{3}{z^5} X^{\dagger}X 
\nonumber \\ 
&- \frac{1}{8g_5^2z}(B_{(L)}^{MN}B_{(L)MN}+B_{(R)}^{MN}B_{(R) MN})\biggr]  \  ,
\end{align}
 where $ D X = \partial X - iB_{(L)}X + iX B_{(R)} $, $ B_{(L,R)} = V \pm A $ and 
$X(x,z) = v(z)U(x,z)/2 $ involves the 
chiral field: $ U(x,z) =  \exp{\left [i \sigma^a \pi^a(x,z)\right ]} $,   with the 
pion field  $\pi^a(x,z)$. The  chiral symmetry is broken by the term 
\mbox{$ v(z) = (m_q z + \sigma z^3) $,}  with $m_q \sim$ quark mass and 
$\sigma $ 
playing the role of quark condensate. 
 The longitudinal
component  of the axial field 
$
 A^a_{\parallel \, M}(x,z) = \partial_M \psi^a(x,z)  \nonumber 
$
gives another pion field  $ \psi^a(x,z) $. 
The model satisfies Gell-Mann--Oakes--Renner relation $m_\pi^2 \sim m_q$. 
In the chiral  limit $m_{q}=0$, it is possible to get the analytic result\cite{DaRold:2005zs,Grigoryan:2007wn}
for $\Psi (z) \equiv \psi (z) -\pi(z)$
\begin{align}
\Psi (z) = 
{z\, \Gamma \left ({2}/{3} \right  )
\left(\frac{\alpha}{2}\right)^{1/3}}
\left[ I_{-1/3}\left(\alpha z^3\right)  -  I_{1/3}\left(\alpha
z^3\right) \frac{I_{2/3}\left(\alpha z^3_0\right)} {I_{-2/3}\left(\alpha z^3_0\right)}\right] \ ,
\end{align}
where $ \alpha = g_5 \sigma/3$. 
  $\Psi (z)$ satisfies $\Psi (0)=1$,  Neumann b.c.  $\Psi '(z_0)= 0$  and 
\begin{align}
f^2_{\pi} =  - \frac{1}{g^2_5}\left(\frac{1}{z}\partial_z \Psi(z) \right)_{z = \epsilon \rightarrow 0}  
\nonumber
\end{align}

\begin{figure}[ht]
\centerline{ \includegraphics[height=2.8cm]{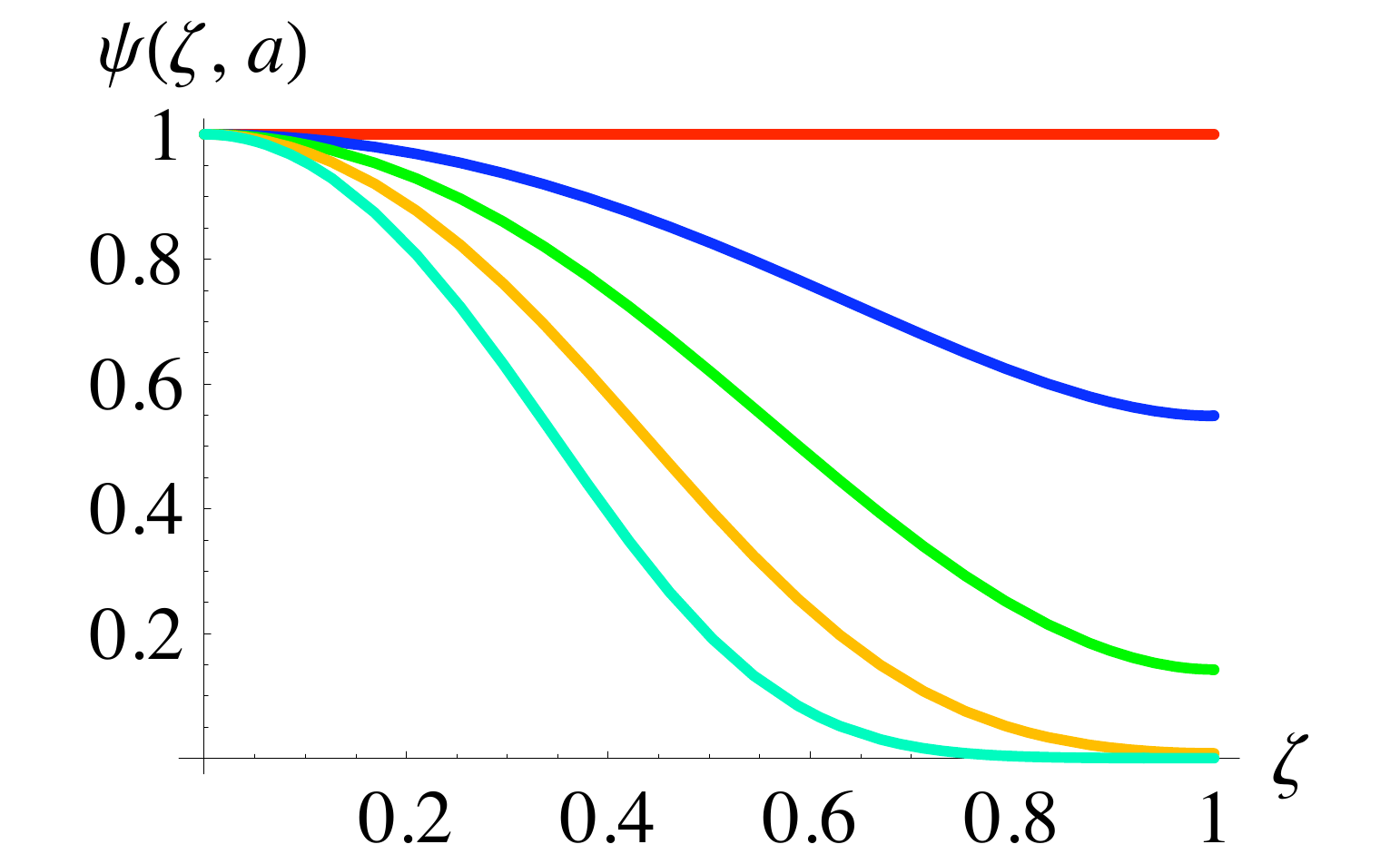}
  \  \includegraphics[height=2.8cm]{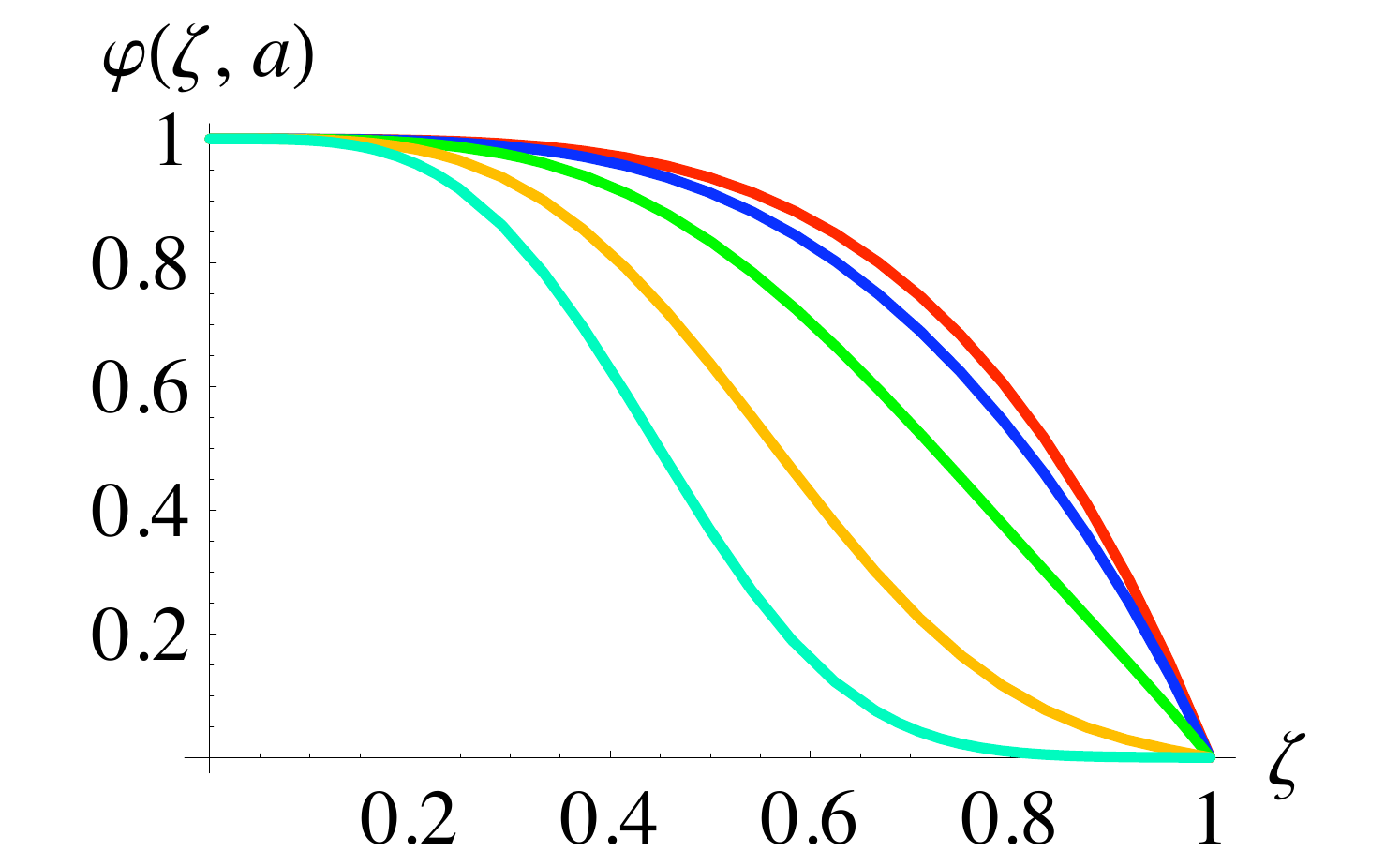}
}
\caption{Pion wave functions $\Psi (z)  \to \psi (\zeta,a)$ and $\Phi (z)  \to \phi (\zeta,a)$ as  functions of 
$\zeta \equiv z/z_{0}$ and 
$a \equiv \alpha z_{0}^{3}$  for $a=0$, 
$a=1$,
$a=2.26$,  $a=5$  and $a=10$. 
}
\end{figure}

 The conjugate wave function  is  given by 
\begin{align}
\Phi(z) = - \frac1{g_5^2 f_\pi^2} \left ( \frac1{z} \, \partial_z \Psi (z) \right)  =
- \frac{2}{s_0}\,  \left ( \frac1{z} \, \partial_z \Psi (z) \right)  \  , 
\end{align}
where   $s_{0} = 4 \pi^{2 } f_{\pi}^{2} \approx 0.67\, $GeV$^{2}$ is the usual 
characteristic scale  for pion. 
The function $\Phi (z)$ satisfies $\Phi (0)=1$ and  Dirichlet b.c.  $\Phi (z_0)= 0$

{\it Pion EM form factor} written 
in terms of $\Psi (z)$  looks like 
\begin{align}
 F_\pi(Q^2) &= \frac{1}{g^2_5 f_\pi^2}\int_{0}^{z_0} {\cal J} (Q,z)\left[\left(\frac{\partial_z\Psi}{z}\right)^2 +
\frac{g^2_5 v^2}{z^4} \Psi^2(z)\right] \, z \, dz \  .
\end{align}
To analyze 
 form factor at large $Q^{2}$, we  write it   in terms of $\Psi (z)$  and $\Phi (z)$:
\begin{align}
 F_\pi(Q^2) &=\int_{0}^{z_0}  {\cal J} (Q,z) \left[{g^2_5 f_\pi^2} \Phi^2 (z)  + \frac{9\alpha^2}
{g^2_5 f^2_{\pi}} \, z^2\, \Psi^2(z) \right] \, z \, dz  \  .
\end{align} 
For large $Q$, only $z \sim 1/Q$ part of  $\Phi^2 (z)$ term works, which gives 
\begin{align}
 F_{\pi}(Q^2) \to \frac{2\, g_{5}^{2}  f_\pi^2 \Phi^2  (0)}{Q^2} =
\frac{4\pi^2 f_\pi^2}{Q^2} \equiv \frac{s_0}{Q^2} \  .
\end{align}
The curve we obtained from the AdS/QCD model (see Ref.[\refcite{Grigoryan:2007wn}] ) goes above 
existing experimental 
data  that  give $Q^2 F_\pi (Q^2)\approx 0.4{\rm GeV}^2$,
  which means  that  the   pion in this model is too small. 

We remind that    pQCD   result\cite{Radyushkin:1977gp,Lepage:1979zb} 
has $2\alpha_s / \pi$ factor  
  \begin{align}
  F_\pi^{{\rm pQCD} } (Q^2) \to \frac{2\alpha_s}{\pi}
\cdot \frac{s_0}{Q^2} 
\sim 0.2 \,  F_\pi^{{\rm AdS/QCD} } (Q^2)
\end{align}
  due to one-gluon exchange.

{\it Anomalous amplitude} of the  $ \gamma^* \gamma^*\pi^0$ transition  is  defined  by 
\begin{eqnarray}
\int
\langle {\pi}, {p}
| T \left\{J^{\mu }_{\rm EM}(x)\,J^{\nu}_{\rm EM}(0)\right\}| 0 \rangle e^{-iq_1 x } d^4 x  \nonumber \\
 = \epsilon^{\mu  \nu \alpha  \beta}
q_{1 \, \alpha} q_{2\, \beta} \, \frac{N_c}{12 \pi^2 f_\pi} \, 
 K_{\gamma^*\gamma^*\pi^0} \left(Q_1^2,Q_2^2 \right )  \  , 
\end{eqnarray}
where $ p = q_1 + q_2 $ and $ q^2_{1,2} = -Q^2_{1,2} $.
Its  value for real photons  is fixed in QCD by axial anomaly: 
$
K_{\gamma^* \gamma^* \pi^0}(0,0) = 1   \ .
$
To  consider this   form  factor, the 
AdS/QCD model  should be extended.   
We need isoscalar fields, which is  achieved by     
gauging $U(2)_L \otimes U(2)_R $ and  introducing the field 
\mbox{$
{\cal B}_{\mu} = t^a   B^a_\mu +   
{\bf 1}
%\mathbb{1} 
\, \frac{{\hat B}_\mu }{2} 
$
}, and we also 
need the Chern-Simons term
\begin{align}
S^{(3)}_{\rm CS}[{\cal B}] =  \frac{N_c}{24\pi^2}\epsilon^{\mu\nu\rho\sigma}&
{\rm Tr} \int d^4 x\, dz  \left(\partial_z
{\cal B}_{\mu}\right)
%\nonumber \\ & \times
\biggl[{\cal F}_{\nu \rho}{\cal B}_{\sigma} + {\cal B}_{\nu}{\cal F}_{\rho \sigma} \biggr]  \  .
\end{align}
The anomalous form factor   conforming to QCD anomaly is  given  by 
\begin{eqnarray}
K (Q_1^2,Q_2^2)  &=&
\Psi (z_0)  {\cal J}(Q_1,z_0) {\cal J}(Q_2,z_0)  \nonumber
	 \\ &-&    \int_0^{z_0}  {\cal J}(Q_1,z) {\cal J}(Q_2,z)\,   \partial_z \Psi (z) \, dz \   . 
\end{eqnarray}
  For large $Q_{1}$ and/or $Q_{2}$ we  may   write 
\begin{eqnarray}
K(Q_1^2,Q_2^2) \simeq  
  \frac{s_0}{2}   \int_0^{z_0}  {\cal J}(Q_1,z)  {\cal J}(Q_2,z) \,  \Phi (z) \, z\, dz  \   .
\end{eqnarray}
If one of the photons is real,  we have 
\begin{eqnarray}
 K  (0,Q^2)   \to \frac{  \Phi (0) s_0}{2 Q^2}
\int_0^\infty d\chi \, \chi^2 \,  K_1 (\chi) =  \frac{ s_0}{Q^2}   \  .
\end{eqnarray}
For  comparison, in pQCD $\gamma^* \gamma \pi^0$  form factor
is  given  by 
$$
 K^{\rm pQCD} (0,Q^2) = \frac{s_0}{3Q^2} \int_0^1 \frac{\varphi_\pi (x)}{x} \, dx
\equiv   \frac{s_0}{3Q^2} \, I^{\varphi} \  .
$$
 The pQCD result  agrees  with AdS/QCD model if  $I^{\varphi} =3$, e.g., for 
\mbox{$\varphi_\pi (x) = 6 x (1-x)$}     (asymptotic DA). 
Our model\cite{Grigoryan:2008cc}   is very close to  Brodsky-Lepage interpolation 
$
 K^{\rm BL} (0,Q^2) = 1/(1+Q^2/s_0) 
$
which goes above CLEO data. However,  next-to-leading 
 pQCD  correction is negative which allows to fit CLEO    data if  one takes 
distribution  amplitudes  with $I^{\varphi} \approx 3$.

 In   case of large and equal   photon virtualities,  the AdS/QCD result is 
\begin{align}
 K(Q^2,Q^2)   \to \frac{  \Phi (0) s_0}{Q^2}
\int_0^\infty d\chi \, \chi^3 \,  [K_1 (\chi)]^2 =  \frac{  s_0}{3Q^2}  \  . 
\end{align}
Note that pQCD result in this   kinematics does not depend on pion DA 
\begin{align}
 K^{\rm pQCD} (Q^2,Q^2) = \frac{s_0}{3} \int_0^1 \frac{\varphi_\pi (x) \, dx}{xQ^2 + (1-x)Q^2} \,
= \frac{s_0}{3Q^2}
\end{align}
 and 
{\it  coincides}  with AdS/QCD model!

For non-equal  large  photon virtualities,  we write 
 $Q_1^2 = (1+\omega)Q^2$ and  $Q_2^2 = (1-\omega)Q^2$. 
The leading-order pQCD  then gives  
\begin{align}
 K^{\rm pQCD} (Q_{1}^2, Q_{2}^2) = %&
\frac{s_0}{3Q^2} \int_0^1 \frac{\varphi_\pi (x) \, dx}{1+\omega (2x-1)}% \nonumber \\
\equiv
%&
\frac{s_0}{3Q^2} \, I^\varphi (\omega)    \  , 
\end{align}
while the AdS/QCD model result reads 
\begin{align}
  K(Q_1^2,Q_2^2) & \to  \frac{  \Phi (0) s_0}{2Q^2} \, \sqrt{1-\omega^2}
\int_0^\infty d\chi \, \chi^3 \,  K_1 ( \chi \sqrt{1+\omega})
 K_1 ( \chi \sqrt{1-\omega})  \nonumber \\
 & =\left ( \frac{s_0}{3Q^2} \right ) \left \{ \frac3{4\omega^3} \, \left [ 2 \omega - (1-\omega^2) \,
\ln \left ( \frac{1- \omega}{1+\omega} \right ) \right ] \right \}  \  .
\end{align}
Note, that the term enclosed in curly brackets 
{\it  coincides}  with  pQCD $ I^\varphi (\omega)$ for $\varphi (x) = 6x (1-x)$.
Indeed, 
using   representation
\begin{equation}
 \chi  K_1 ( \chi ) = \int_0^\infty e^{-\chi^2/4u -u} \, du \  ,
\end{equation}
 and  integrating  over $\chi$ we  get
\begin{eqnarray}
\label{KQ1Q2}
K(Q_1^2, Q_2^2)  \to  \frac{  s_0}{Q^2} \,
\int_0^\infty  \int_0^\infty \frac{u_1 u_2 \,e^{-u_1-u_2}  du_1 du_2}{u_2 (1+\omega)  +u_1(1-\omega)}
  \,  .
\end{eqnarray}
 Changing  $u_2=x\lambda$, $u_1= (1-x) \lambda$
and integrating   over $\lambda$ gives 
\begin{align}
 K(Q_1^2, Q_2^2)  \to  &
\frac{s_0}{3Q^2} \int_0^1 \frac{6 \, x (1-x) \, dx}{1+\omega (2x-1)} \ .
\end{align}

{\it Comment on ``Light-Front Holography''.} The AdS/CFT form factor   expression  (\ref{AdSFF}) 
 has structure similar to that of DY light-front formula (\ref{DY}), especially 
when the latter is written in terms of the impact parameter  space w.f. $\widetilde \Psi (x,{\bf b}_\perp)$.
Brodsky and de Teramond\cite{Brodsky:2006uqa}  noticed that, 
identifying $z$ with $|{\bf b}_\perp|\sqrt{x (1-x)}$ and  taking 
a special form  of the light-front w.f. 
\begin{align}
\widetilde \Psi (x, {\bf b}_\perp) = \frac1{\sqrt{2\pi}}  \, 
\frac{\Phi ( |{\bf b}_\perp|\sqrt{x (1-x)})}{ {\bf b}_\perp^2 \sqrt{x (1-x)}} \  ,
\end{align}
one can  convert  the 3D DY formula (\ref{DY})  into the 
1D AdS/CFT  integral   (\ref{AdSFF}). 
This  observation is the basis of the ``Light-Front Holography''  
approach\cite{deTeramond:2008ht}.
However, it is  easy to check   that if one would calculate the meson  couplings $f_n$ 
(\ref{fn}),  (\ref{fnS}) 
from the  light-front w.f. fixed by this ansatz, the results would have an 
extra $\sqrt{6} \pi/8$ 
factor (see Eqs.(88),(89) of Ref.[\refcite{Brodsky:2007hb}]) compared to exact  AdS/QCD results
 (\ref{fn}),  (\ref{fnS}). 
Furthermore, this  ansatz gives $8 \sqrt{x(1-x)}/\pi$  for meson distribution amplitude,
while we demonstrated above that AdS/QCD results  
for \mbox{$\gamma^*\gamma^* \to \pi^0$}  form factor 
correspond  to asymptotic $6x(1-x)$ distribution amplitude.
In general, the light-front  holography ansatz\cite{Brodsky:2006uqa}  
is  not consistent with AdS/QCD  for any observable that depends   linearly  on the w.f. 
(rather than bilinearly as in DY formula).

\section{Summary}

  Summarizing,  we established that meson 
form factors in  AdS/QCD are given by  formulas
similar  to those in quantum  mechanics. 
 For large $Q$, there is only one mechanism $z \sim 1/Q$. 
For  vector mesons, 
the  leading   ({\it LL}) IMF form factor  ${\cal F} (Q^{2})$ 
 indeed behaves like $1/Q^{2}$ for large $Q^{2}$. 
In soft-wall model, ${\cal F} (Q^{2})$ demonstrates exact $\rho$-dominance. 
 For pion, large-$Q^{2}$ asymptotics is 
$s_{0}/Q^{2}$ vs. pQCD  result $(2 \alpha_{s}/\pi) s_{0}/Q^{2}$.
 We included the anomalous amplitude into the AdS/QCD analysis,
extending it  to $U(2)_{L}\otimes U(2)_{R}$ and adding the Chern-Simons term. 
 Fixing normalization by conforming to QCD anomaly, we observed that 
 large-$Q^{2}$ behavior coincides  then with pQCD calculations
for asymptotic pion DA, the result contradicting the  claim of ``light-front holography''
approach that meson distribution amplitude  is  given  by $8\sqrt{x(1-x)}/\pi$.
In conclusion,  AdS/QCD is an  instructive model   for what 
may happen  with form factors  in real-world  QCD. 

 \section*{Acknowledgements}

I am very grateful to Organizers for invitation to Workshop honoring 
 60th anniversary of M.~Shifman and their hospitality. Happy birthday, Misha!

I thank H.R. Grigoryan for  collaboration on  the studies of form factors in AdS/QCD. 

This paper is authored by Jefferson Science Associates, LLC under U.S. DOE Contract No. DE-AC05-06OR23177.
The United States Government 
retains and the publisher, by accepting the article for publication, 
acknowledges that the United States Government retains a non-exclusive, 
paid-up, irrevocable, world wide license to publish or reproduce 
the published form of this manuscript, or allow others to do so, 
for United States Government purposes.

\end{document}